\documentclass[aip,reprint,twocolumn,groupedaddress,apl]{revtex4-1} 

\usepackage{graphicx}
\usepackage{textcomp}
\usepackage{amsmath}
\usepackage{natbib}
\usepackage{dcolumn}
\usepackage{bm}

\begin{document}
\title{Planar Superconducting Whispering Gallery Mode Resonators  }
\author{Z.K. Minev}
\author{I.M. Pop}
\author{M.H. Devoret}
\affiliation{Department of Applied Physics, Yale University, New Haven, Connecticut 06511, USA}

\date{\today}

\begin{abstract}
We introduce a microwave circuit architecture for quantum signal processing combining design principles borrowed from high-Q 3D resonators in the quantum regime and from planar structures fabricated with standard lithography. 
The resulting `2.5D' whispering-gallery mode resonators store 98\% of their energy in vacuum.
We have measured internal quality factors above 3 million at the single photon level and have used the device as a materials characterization platform to place an upper bound on the surface resistance of thin film aluminum of less than $250$n$\Omega$.
\end{abstract}

\maketitle

Low loss microwave structures form the basis for emerging superconducting quantum-based technologies. 
Of key interest are high quality resonators for superconducting detectors,\cite{Day2003} microwave nanomechanics \cite{Regal2008,Cicak2010} and circuit quantum electrodynamics (cQED), where they are used to perform quantum gates\cite{Kirchmair2013} or to store quantum information\cite{Mariantoni2011}.
Single photon quality factors (Q's) of planar (2D) resonators approach $2\times10^6$ by careful geometric and material engineering.\cite{Megrant2012}
Experiments show that resonators with larger features which dilute the electromagnetic energy density in dielectrics and at interfaces tend to improve the Q's.\cite{Gao200815,Barends2010,Geerlings2012,Megrant2012,Zmuidzinas2012}
Recently, this trend was extended to the extreme in 3D resonators\cite{Paik2011} which can now exceed power independent Q's of $500\times10^6$ at single photon powers.\cite{Reagor2013} 
A similar approach which aims at minimizing energy densities in dielectrics is the development of vacuum-gap-based capacitors; planar resonators measured with these capacitors reach internal Q's of $\approx 0.15\times10^6$.\cite{Cicak2010}   
Is it possible to marry the advantages of 3D microwave structures with those of planar fabricated circuits for superconducting quantum-based technologies?

Inspired by the optics community,\cite{Vahala2003,Ilchenko2006} we introduce a superconducting whispering-gallery mode (WGM) resonator assembled from two wafers patterned with standard lithography techniques (see Fig.~1).
With this strategy, we confine $98\%$ of the WGM energy in lossless vacuum and measure power-independent single-photon lifetimes $T_1 = 180\mu$s ($Q_{int} = 3.4\times10^6$) at 15mK.
Furthermore, this design allows us to selectively probe the metal-air interface of thin film Aluminum (Al), for which we place an upper bound on the surface resistance $R_S < 250$n$\Omega$. 
Combining the advantages of 3D resonators and structures with those of planar circuits and qubits remains a standing challenge in the field.

The WGM resonator can be constructed from the textbook example of the ideal parallel plate transmission line with periodic boundary conditions (Fig.~1-b).\cite{Pozar} 
This transmission line confines the electric and magnetic fields in the vacuum between the plates. 
For a real structure, the periodic boundary condition can be realized by wrapping the transmission line into a ring-like structure.\cite{Eriksson2001}
By breaking the circular symmetry of the rings we obtain a non-degenerate pair of standing modes, one 
parallel (${\parallel}$) and one perpendicular (${\perp}$) to the symmetry axis (Fig.~1-a).
The horizontal symmetry plane allows for a separation of modes into common and differential ones. 
The common modes ($C$) have mirror charges on the upper and lower plates, while the differential ones ($D$) have opposite charges (Fig.~1-c,d).

\begin{figure}[b]
\begin{centering}
\includegraphics[width=3.375in]{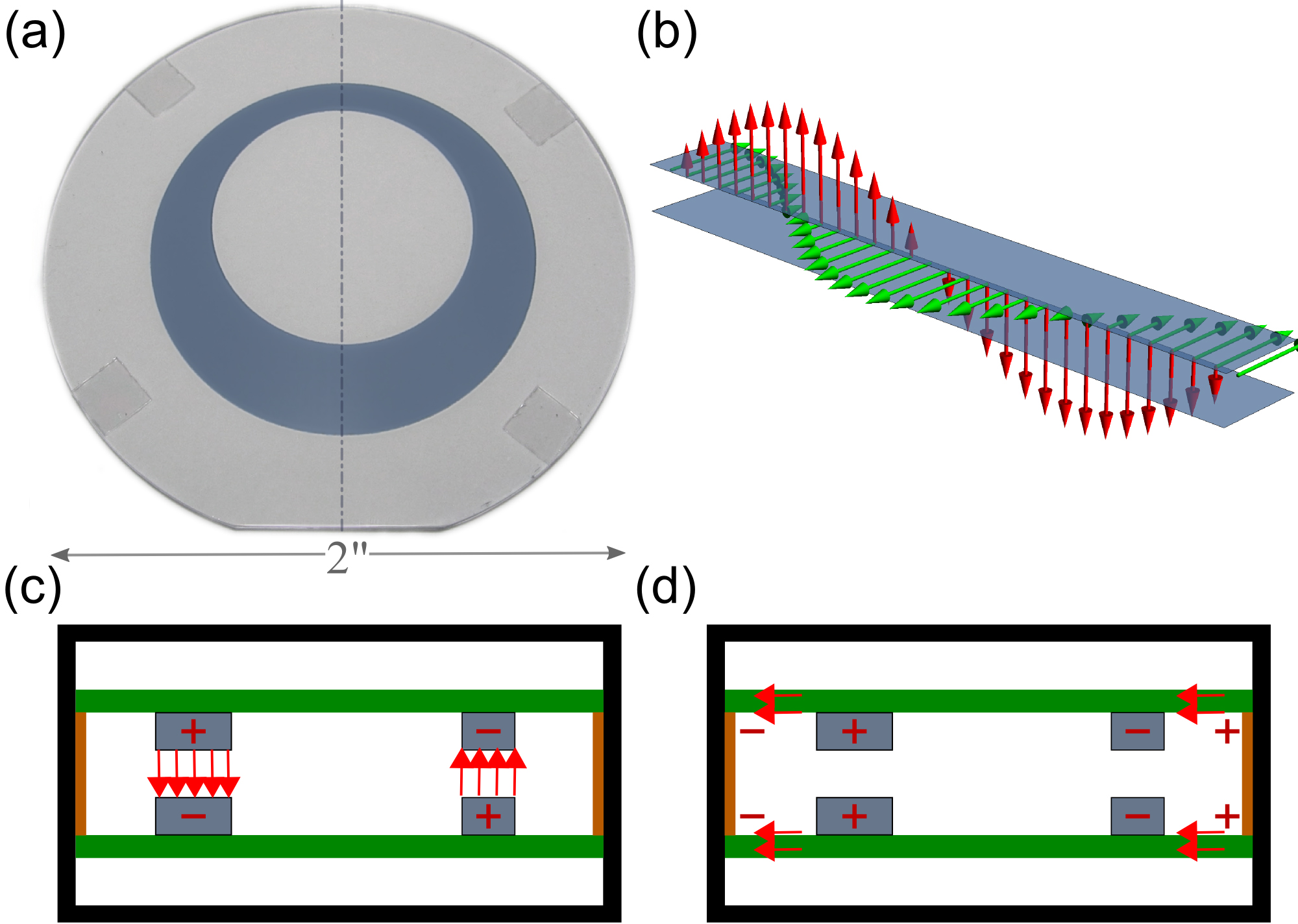}
\caption{
a) Picture of WGM ring resonator comprised of two thin film rings deposited on different sapphire wafers separated by $200\mu$m sapphire spacers (small transparent rectangles). 
The elements are bonded together with PMMA. 
The dashed vertical line shows the symmetry plane which defines the parallel (${\parallel}$) and perpendicular (${\perp}$) mode orientation.
b) The WGM resonator can also be thought of as a transmission line closing on itself. 
Electric (red) and magnetic (green) field pattern of lowest standing mode of a parallel plate transmission line with periodic boundary conditions.
c) A not-to-scale cross-sectional representation of the differential D${^\parallel}$ mode along the dashed line in (a). 
The charges on the top and bottom ring are different and the electric field lines (red) span the vacuum between the rings. 
HFSS simulations estimate 98\% of the field's energy is confined in the vacuum between the two thin film rings (blue). 
The orange rectangles caricature the sapphire separators. 
d) The common modes ${C^\parallel}$ and ${C^\perp}$ store only ~55\% of their energy in vacuum, while the rest is housed in the sapphire dielectric (green).}
\label{fig:SampleHolderAndModes}
\end{centering}
\end{figure}

\begin{figure*}
\includegraphics[width=6.9in]{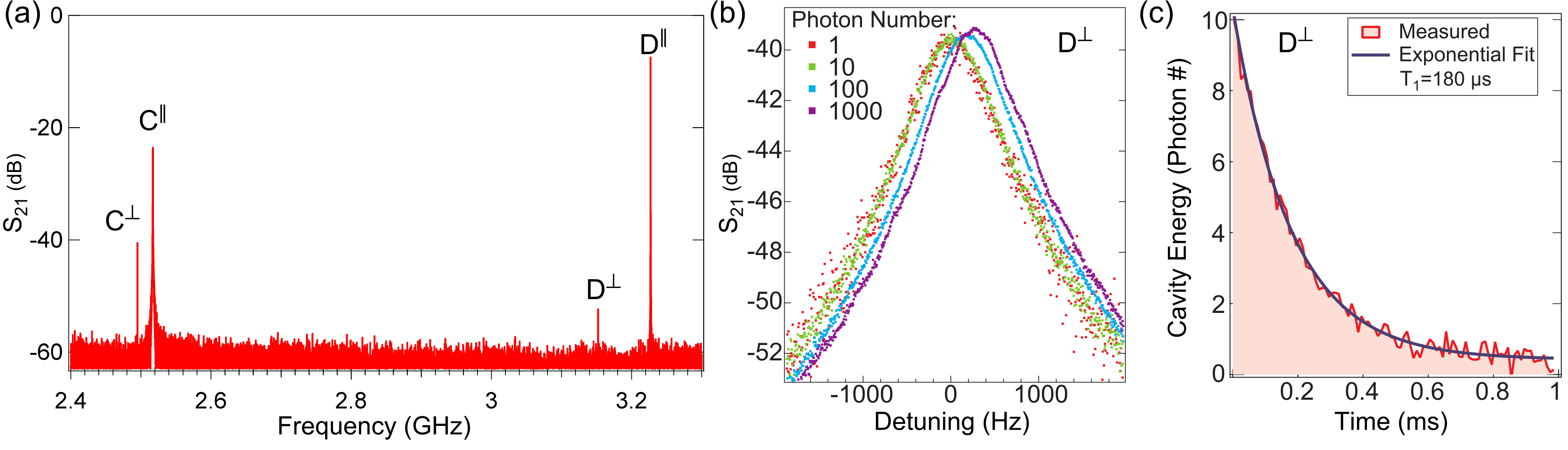}
\caption{
a) $S_{21}$ spectroscopy measurement at 15mK revealing the 4 lowest modes of the resonator.
b) The highest Q D$^\perp$ mode has a power independent $Q=3.37\times10^6$ (or $T_2^* = 340\mu$s) at the single photon level. The slight upward shift in frequency with increasing power can be attributed to residual vortices in the rings, trapped during cool-down, which become mobile at higher driver powers.
c) A heterodyne time domain ring-down measurement of the D$^\perp$ mode in (b) is fitted with a single exponential with a decay time $T_1 = 180\mu$s. The decay time inferred from the line width and measured decay time agree, indicating $T_2 \approx 2T_1$.}
\label{fig:S21QandHetero}
\end{figure*}

We pattern the rings on 2" sapphire wafers using a PMMA/MMA resist bilayer and electron beam exposure. 
A $300$nm Al film is deposited using electron beam deposition followed by liftoff.
The two wafers are separated using four 5x2x0.2 mm sapphire spacers placed on the edges of the wafers in locations of minimal mode participation. 
The structure is bonded using PMMA and the resulting resonator (Fig.~1-a) is placed inside a two-piece pure Al 5N5 (99.9995\%) sample holder as depicted in Fig.~1-c,d. 
Non-magnetic pins penetrate the top lid directly above the thinnest and thickest part of the rings, serving as a drive and readout port respectively. 

The metallic sample holder perturbs the boundary conditions and inevitably introduces new modes. 
However, HFSS finite element simulations confirm that the $D$ modes remain intact with 98\% of their energy stored in the vacuum (Fig.~1-c), while the $C$ modes are strongly perturbed and 45\% of their energy is housed in the sapphire wafers.

The resonator samples are mounted inside a Cryoperm shield and are thermally anchored to the mixing chamber plate (15mK) of a dilution unit.
The amplification chain consists of two Pamtech isolators and a 12 GHz K\&L multi-section low-pass filter followed by a HEMT amplifier. 

In Fig.~2-a we present a characteristic $S_{21}$ transmission measurement, where the 0dB correspond to a through cable replacing the sample using a low temperature switch. 
The common modes (${C^\parallel}$, ${C^\perp}$) are approximately $0.5$ GHz lower than the differential modes (${D^\parallel}$, ${D^\perp}$) due to the large participation of the high permittivity sapphire. 
The asymmetry induced splitting between the ${D^\parallel}$ and ${D^\perp}$ modes is $75$ MHz.
The noise floor of the measurement is set by the averaging time of $100$ ms per point.
The measured frequencies of the 4 modes agree within 2\% with numerical simulations.

In Fig.~2-b we present $S_{21}$ measurements of the ${D^\perp}$  mode, and extract $Q = 3.37\times10^6 \pm 0.15$ corresponding to $T_2^* = 340 \pm 15 \mu$s at the single photon level.  
The $-40$ dB insertion loss of the ${D^\perp}$ mode indicates that it is undercoupled and its Q is dominated by internal losses. 
The fitted coupling quality factor is $Q_c = 300\times10^6$.\cite{Pozar}
The average photon number inside the resonator is given by $\bar{n}=P_{in}Q/\hbar\omega^{2}$, where $P_{in}$ is the power delivered to the cavity and $\omega$ is the mode's angular frequency on resonance.\cite{Sears2012,Sage2011}
The measured Q varies less than 3\% for photon number excitations ranging from $1$ to $10^3$.
At high photon numbers ($\overline{n} > 100$) we observe a small upward frequency shift of $\approx 250$Hz, which may be explained by trapped vortices due to residual magnetic fields. 
The measured quality factor is reproducible with thermal cycling.
\begin{figure}[b] 
\includegraphics[width=3.375in]{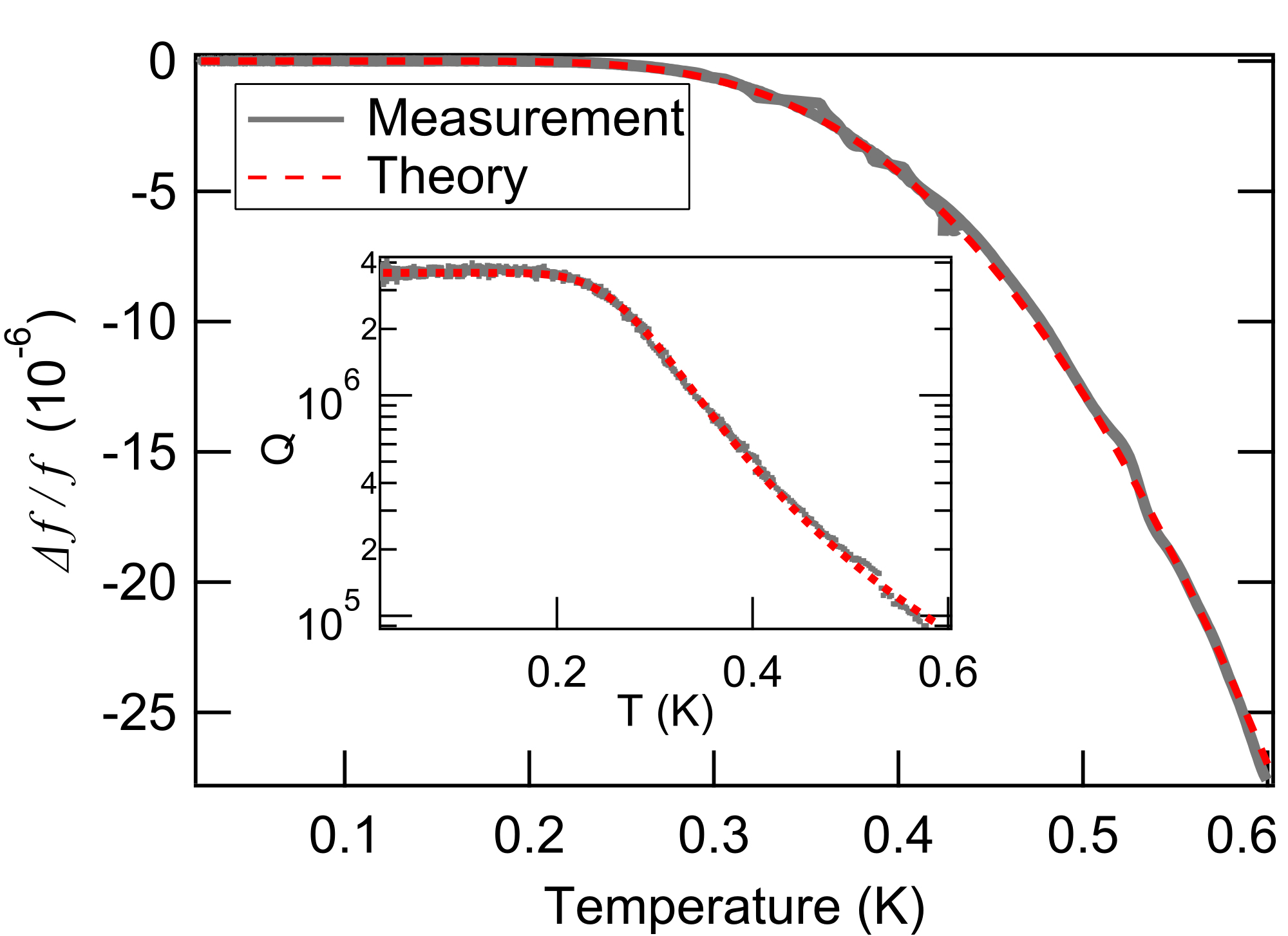}
\caption{
D$^\perp$  mode resonance frequency shift as a function of temperature. The Mattis-Bardeen theory fit (red dashes) provides an upper bound on the surface resistance of thin film Al of less than $250$n$\Omega$. 
Inset shows corresponding quality factor dependence on temperature.}
\label{fig:TempAndSapphireQ}
\end{figure}
To measure the energy decay time $T_1$ of the resonator, we use an RF heterodyne setup to monitor its average output power versus time during ring-down. 
In Fig.~2-c we present a typical ring-down for the ${D^\perp}$ mode and show an exponential fit with decay time $T_1 = 180\mu$s. 
We note that the characteristic decay times measured using phase-sensitive (Fig.~2-b) and phase-insensitive (Fig.~2-c) heterodyne techniques agree ($\frac{T_2^*-2T_1}{T_2^{*}}<6\%$), thus placing a lower bound for $T_\phi > 10 ~T_1$.

The WGM resonator design confines the energy of the differential modes in the vacuum between the thin-film Al tracks, thus limiting losses to the Al-vacuum interface.
We use this feature to study the surface quality of the superconducting thin-film. 
The film surface impedance can be assessed by measuring the fractional frequency shift of ${D^\perp}$ versus temperature which depends on the kinetic inductance fraction ($\alpha$), the material dependent effective penetration depth ($\lambda$) and the real and imaginary parts of the surface impedance $R_s$ and  $X_{s} = \omega \mu_0 \lambda+\delta X_{s}(T)$:\cite{Reagor2013,Turneaure1991,Gao2008} 
\begin{equation}\label{eq:surfImp}
\frac{1}{Q}+2j\frac{\delta f}{f} = \frac{\alpha}{\omega\mu_{0}\lambda} (R_{s}+j\delta X_{s}),
\end{equation}
where $\mu_0$ is the permeability of vacuum.
From the experimental data shown in Fig.~3 we extract $\alpha = 1.4\times 10^{-3}$, which agrees with numerical simulations and is consistent across multiple samples and cool downs. 
This kinetic inductance fraction is in between typical values for planar resonators ($10^{-2} < \alpha < 1$) and 3D cavities ($10^{-5} < \alpha < 10^{-6}$). \cite{Day2003,Leduc2010,Reagor2013}
From the real part of Eq. (\ref{eq:surfImp}) we find an upper bound on the surface resistance of thin-film Al of $R_s < 250$n$\Omega$ corresponding to a surface quality factor $Q_s =\frac{X_S}{R_S} > 4,800$ at the single-photon level.
The power independent $T_1$ decay rate (Fig.~2-c) and the monotonic Q and $\delta f$ (Fig.~3) variation with temperature indicate that the WGR modes are not influenced by TLS. \cite{Martinis2005} 

For the under-coupled  ${C^\perp}$ mode, we measure a power independent single-photon $Q = 2.1\times 10^6$.
From the $44\%$ sapphire participation ratio of this mode, we place a bound on the single photon loss tangent of sapphire of $\tan \delta < 10^{-6}$, which is comparable with values extracted from superconducting qubit energy relaxation times. \cite{Paik2011}

In conclusion, we have presented a direction for constructing superconducting high-Q resonators compatible with both cQED experiments and wafer level fabrication.
This simple and robust architecture can integrate high-Q planar resonators with qubits and control lines using standard lithography and flip-chip techniques. 

We thank Teresa Brecht, Kurtis Geerlings, Leonid Glazman, Matt Reagor, Rob Schoelkopf and Kyle Serniak for valuable discussions. 
Simulations were aided by Dominic Kwok. 
This research was supported by IARPA under Grant No. W911NF-09-1-0369, ARO under Grant No. W911NF-09-1-0514 and NSF under Grant No. DMR-1006060.
Facilities use was supported by YINQE and NSF MRSEC DMR 1119826.

\bibliographystyle{apsrev4}

\end{document}